\documentclass[a4paper,fleqn,usenatbib]{mnras}

\usepackage{txfonts}

\usepackage{graphicx}	

\newcommand{\kms}{km~s$^{-1}$}
\newcommand{\msun}{$M_{\odot}$}

\newcommand{\mh}{$M_{H_2}$}
\newcommand{\mhi}{$M_{HI}$}
\newcommand{\mstar}{$M_{\ast}$}
\newcommand{\must}{$\mu_{\ast}$}

\newcommand{\rmol}{$R_{mol}$}
\newcommand{\tdep}{$t_{dep}({\rm H_2})$}

\newcommand{\fgas}{$f_{\rm H_2}$}
\newcommand{\fhi}{$f_{\rm HI}$}
\newcommand{\xco}{$\alpha_{CO}$}
\newcommand{\ms}{SFR-$M_{\ast}$}
\newcommand{\sfruv}{SFR$_{\rm UV}$}

\newcommand{\ntothi}{760}

\title[Gas along and across the main sequence]{Molecular and atomic gas along and across the main sequence of star-forming galaxies}

\author[A. Saintonge et al.]{Amelie Saintonge,$^{1}$\thanks{E-mail: a.saintonge@ucl.ac.uk}
Barbara Catinella,$^{2}$
Luca Cortese,$^{2}$
Reinhard Genzel,$^{3}$ 
\newauthor Riccardo Giovanelli,$^{4}$
Martha P. Haynes,$^{4}$
Steven Janowiecki,$^{2}$
\newauthor Carsten Kramer,$^{5}$
Katharina A. Lutz,$^{6,7}$
David Schiminovich,$^{8}$
\newauthor Linda J. Tacconi,$^{3}$
Stijn Wuyts$^{9}$
and Gioacchino Accurso$^{1}$
\\
$^{1}$Department of Physics and Astronomy, University College London, Gower Street, London, WC1E 6BT, UK\\
$^{2}$International Centre for Radio Astronomy Research (ICRAR), The University of Western Australia, Perth, WA 6009, Australia\\
$^{3}$Max-Planck Institut fur extraterrestrische Physik, 85741 Garching, Germany\\
$^{4}$Cornell Center for Astrophysics and Planetary Science, Space Sciences Building, Cornell University, Ithaca, NY 14853\\
$^{5}$Instituto Radioastronom\'{i}a Milim\'{e}trica, Av. Divina Pastora 7, Nucleo Central, 18012 Granada, Spain\\
$^{6}$Centre for Astrophysics and Supercomputing, Swinburne University of Technology, PO Box 218, Hawthorn, VIC 3122, Australia\\
$^{7}$Australia Telescope National Facility, CSIRO, P.O. Box 76, Epping, NSW 1710, Australia\\
$^{8}$Department of Astronomy, Columbia University, New York, NY 10027, USA\\
$^{9}$Department of Physics, University of Bath, Claverton Down, Bath, BA2 7AY, UK
}

\date{Accepted XXX. Received YYY; in original form ZZZ}

\pubyear{2016}

\begin{document}
\label{firstpage}
\pagerange{\pageref{firstpage}--\pageref{lastpage}}
\maketitle

\begin{abstract}
We use spectra from the ALFALFA, GASS and COLD GASS surveys to quantify variations in the mean atomic and molecular gas mass fractions throughout the \ms\ plane and along the main sequence (MS) of star-forming galaxies.   Although galaxies well below the MS tend to be undetected in the Arecibo and IRAM observations, reliable mean atomic and molecular gas fractions can be obtained through a spectral stacking technique.  We find that the position of galaxies in the \ms\ plane can be explained mostly by their global cold gas reservoirs as observed in the HI line, with in addition systematic variations in the molecular-to-atomic ratio and star formation efficiency.  When looking at galaxies within $\pm0.4$ dex of the MS, we find that as stellar mass increases, both atomic and molecular gas mass fractions decrease, stellar bulges become more prominent, and the mean stellar ages increase.  Both star formation efficiency and molecular-to-atomic ratios vary little for massive main sequence galaxies, indicating that the flattening of the MS is due to the global decrease of the cold gas reservoirs of galaxies rather than to bottlenecks in the process of converting cold atomic gas to stars. 
\end{abstract}


\section{Introduction}
\label{intro}

The current galaxy evolution framework gives central stage to the cycling of gas in and out of galaxies, and the efficiency of the star formation process out of the gas that cools and settles into galactic discs.  These elements are responsible for regulating star formation in galaxies, much more so than, for example, merger-driven starbursts. Simple numerical and analytical models that focus on the gas reservoir of galaxies, and how it is replenished by inflows and depleted by star formation and outflows \citep[e.g.][]{frenkwhite91,bouche10,dave11,dave12,krumholz12,lilly13}, are well supported by observations of the redshift evolution of the molecular gas contents of galaxies \citep{tacconi10,tacconi13,magdis12a,saintonge13,genzel15}. These studies show that the redshift evolution of the specific star formation rate (SSFR) can be explained simply by the changes in the molecular gas contents of galaxies, and by the mean star formation efficiency that increases slightly with redshift. 

Other studies have looked at variations in the gas contents and star formation efficiency of galaxies as a function of SSFR, or distance from the star formation main sequence.  At all redshifts up to $z\sim2$, it has been established that galaxies well above the main sequence have star formation efficiencies enhanced by an order of magnitude \citep[e.g.][]{gao04,genzel10,daddi10}, and that almost all of these most extreme systems are major mergers \citep{sanders96,veilleux02,engel10}.  Further studies have shown that star formation efficiency is not a bimodal quantity, but rather varies smoothly as a function of SSFR, or distance from the MS \citep{COLDGASS2}.  The molecular gas contents of galaxies also varies with distance from the MS, indicating that the high SSFRs of starbursting galaxies are caused both by enhanced molecular gas mass fractions and increased star formation efficiencies, while the reverse effect is observed in bulge-dominated, below-MS galaxies \citep{saintonge12,genzel15}. 
 
In this paper, we push these investigations further in two main respects.  Firstly, we focus on a local sample, which gives us access not only to the molecular gas, but also to the cold atomic gas contents of galaxies through HI 21cm observations.  By studying both atomic and molecular gas fractions, we can identify if the conversion of atomic to molecular gas is in any region of the SFR-\mstar\ plane a bottleneck in the star formation process.  Secondly, instead of restricting our study to gas-rich main sequence galaxies, we take advantage of the size and completeness of the GASS sample and make use of a spectral stacking technique to measure mean molecular and atomic gas fractions in the entire SFR-\mstar\ plane, including galaxies with SFRs more than two orders of magnitude lower than main sequence galaxies of the same mass.  

All rest-frame and derived quantities assume a \citet{chabrier03} IMF, and a cosmology with $H_0=70$\kms\ Mpc$^{-1}$, $\Omega_m=0.3$ and $\Omega_{\Lambda}=0.7$.

\section{Data}
\label{sample}

\begin{figure}
\includegraphics[width=\columnwidth]{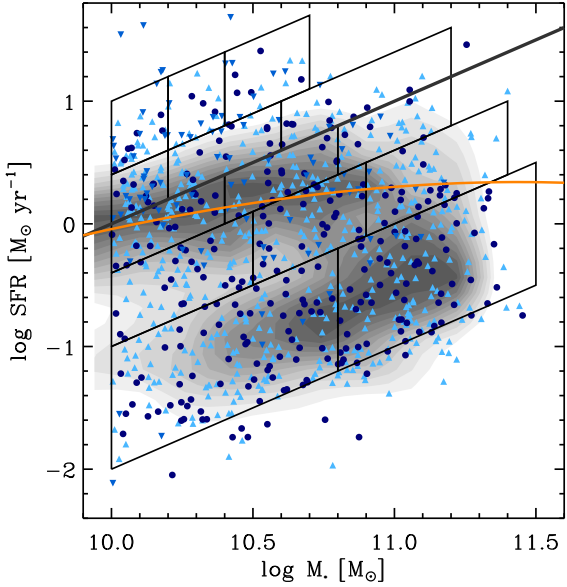}
\caption{Sample distribution in the SFR-\mstar\ place.  The grayscale contours show the underlying distribution of the 12006 SDSS galaxies that form the complete parent sample from which the GASS targets were randomly selected. Data points represent galaxies that have been observed by Arecibo and/or IRAM: filled dark blue circles have both HI and CO observations, light blue triangles only HI, and down-pointing triangles only CO observations.  The black thick solid line represent a constant SSFR value of 0.1 Gyr$^{-1}$, the characteristic value for star forming galaxies of \mstar$=10^{10}$\msun\ at $z=0$, and the orange line follows the star formation main sequence (as parametrised in Eq. \ref{MSeq}).  The boxes represent the regions of the plane that are stacked together to derive mean gas fractions as explained in \S \ref{stacking}. \label{figsample}}
\end{figure}

\subsection{Sample selection and Arecibo/IRAM observations}

The GASS sample was selected from the SDSS DR7 spectroscopic sample based purely on redshift ($0.025<z<0.050$) and stellar mass (\mstar$>10^{10}$\msun), and is therefore representative of the entire population of local, massive galaxies.   

A detailed description of the GASS sample selection and data products is given in \citet{GASS1,GASS6,GASS8}.   Out of a parent sample of 12006 galaxies matching these selection criteria, \ntothi\ objects were randomly selected for HI observations at the Arecibo Observatory. The most gas-rich galaxies in the sample are already detected in the ALFALFA survey \citep{alfalfa1,haynes11} and therefore not reobserved. To increase the number of galaxies with high SSFR ($\log {\rm SSFR}>-9.6$), we supplement the GASS sample with galaxies from the SDSS parent sample with spectra available from ALFALFA.  To avoid any biases, we include both high-SSFR galaxies with and without ALFALFA HI detections by extracting spectra from the data cubes centered on the SDSS optical positions. 

A randomly selected subset of 350 galaxies out of the GASS sample (and of its parent super-sample) was further targeted for molecular gas observations.  The COLD GASS survey used the IRAM 30-m with the EMIR receiver to measure CO(1-0) line emission in these galaxies, as described in \citet{COLDGASS1}.  The data were reduced using standard techniques, including aperture corrections to account for the beam size of the telescope at the observed frequency.  Total molecular gas masses are extrapolated from the aperture-corrected CO(1-0) line fluxes using a standard Galactic conversion factor, \xco$=4.35$ \msun (K \kms\ pc$^{2}$)$^{-1}$, except for 12 galaxies all with high SSFR identified through their high IRAS 60$\mu$m/100$\mu$m ratios as requiring a lower conversion factor \citep[\xco$=1.00$ \msun (K \kms\ pc$^{2}$)$^{-1}$, the value typically used for major mergers such as the $z=0$ ULIRGS;][]{solomon97} .  Arguments as to why most of the COLD GASS galaxies are in a regime where a constant Galactic \xco\ value is adequate are presented elsewhere \citep{COLDGASS1,COLDGASS2,saintonge12}.

The most important features of the sample in the context of this study are (1) the lack of preselection against anything other than stellar mass and redshift, and (2) a consistent observing strategy at Arecibo and IRAM allowing us to place stringent upper limits on the gas fractions even when the spectral lines are not detected.  As a result, the sample fully probes the entire \ms\ plane, and even in the absence of individual detections for passive galaxies, meaningful information about the mean gas content of these galaxies can nonetheless be retrieved.  Figure \ref{figsample} shows the distribution of galaxies in the GASS/COLD GASS samples in the SFR-mass plane, highlighting the uniform coverage of this parameter space.  Because the underlying parent sample is complete and well-defined from SDSS (grayscale contours in Fig. \ref{figsample}), unbiased scaling relation or global measurements can be derived by applying a simple weighting scheme \citep{COLDGASS1}. 

\subsection{Stellar masses and star formation rates}

Additional data products necessary for this study are derived using the public SDSS, {\it GALEX} and {\it WISE} imaging surveys.  Stellar masses derived based on the SDSS photometry method of \citet{salim07} are retrieved from the MPA/JHU catalog.  Structural parameters such as stellar mass surface density, \must, and concentration index, $C\equiv R_{90}/R_{50}$, are also derived using SDSS data products \citep[details in][]{GASS1,COLDGASS1}. 

Only the star formation rates used in this work differ from the data products used in previous GASS and COLD GASS studies.  We make use of the public {\it WISE} and {\it GALEX} imaging to derive total SFRs by adding the UV and IR contributions.   We calculate \sfruv\ as 
\begin{equation}
{\rm SFR}_{UV}=6.84\times10^{-29} L_{NUV}
\end{equation}
where $L_{NUV}$ is the luminosity in the {\it GALEX} NUV band using the photometric measurements derived for the purposes of the GASS survey, uncorrected for dust extinction \citep{wang10,GASS1}.  This calibration of the conversion between near-UV flux and SFR was done assuming a Kroupa IMF and a continuous recent star formation history \citep{salim07,schiminovich07}.  We finally apply a 6\% correction factor following \citet{madaudickinson14} to make \sfruv\ consistent with a Chabrier IMF.  

To account for the dust-obscured star formation, we make use of the {\it AllWISE} Atlas images. After masking nearby objects and performing background subtraction, Sextractor \citep{bertin96} is used to derive accurate photometry.  The Kron magnitudes produced by Sextractor are converted from the {\it WISE} Vega system to AB magnitudes using the magnitude offsets prescribed in the {\it WISE} Data Handbook, and corrected for stellar contamination using the  {\it WISE} 3.4$\mu$m photometry following \citet{ciesla14}.  The corrected 12$\mu$m and 22$\mu$m luminosities can then be used to compute dust-obscured SFRs using well-calibrated prescriptions such as that from \citet{jarrett13}:
\begin{eqnarray}
{\rm SFR}_{IR,12} &=& 4.91\times10^{-10} L_{12\mu m}\\ \nonumber
{\rm SFR}_{IR,22} &=& 7.50\times10^{-10} L_{22\mu m}\;. 
\label{SFR22}
\end{eqnarray}

Of the four {\it WISE} bands, the 22$\mu$m is the optimal choice to compute SFR$_{IR}$, but given the relative shallowness of these observations, only 51\% of all COLD GASS galaxies are detected with $S/N>3$ (the 22$\mu$m detection fraction increases to 82\% if considering only the CO-detected galaxies).  The detection fraction is significantly higher at 12$\mu$m (85\% for all COLD GASS galaxies, 93\% for the CO-detected subset). We therefore use $SFR_{tot}=SFR_{IR,22}+SFR_{UV}$ when available, and $SFR_{IR,12}+SFR_{UV}$ otherwise.  Comparing the values of $SFR_{tot}$ with the measurements from optical-UV SED fitting \citep{COLDGASS2} reveals no systematic offset and a 0.3 dex scatter, in complete agreement with \citet{huang14} who also derived SFRs based on {\it GALEX} and {\it WISE} photometry for COLD GASS galaxies.  In this paper, we use $SFR_{tot}$ for all galaxies with $S/N>3$ in both {\it GALEX} NUV and either of the {\it WISE} bands, and $SFR_{SED}$ otherwise; such calibrated SFR ladders have been shown to be reliable when measuring SFRs across large galaxy samples including both passive and actively star-forming objects \citep[e.g.][]{wuyts11sfr}.

\section{Results}
\label{results}

\subsection{Atomic and molecular gas in the \ms\ plane}
\label{stacking}

\begin{figure*}
\includegraphics[width=\textwidth]{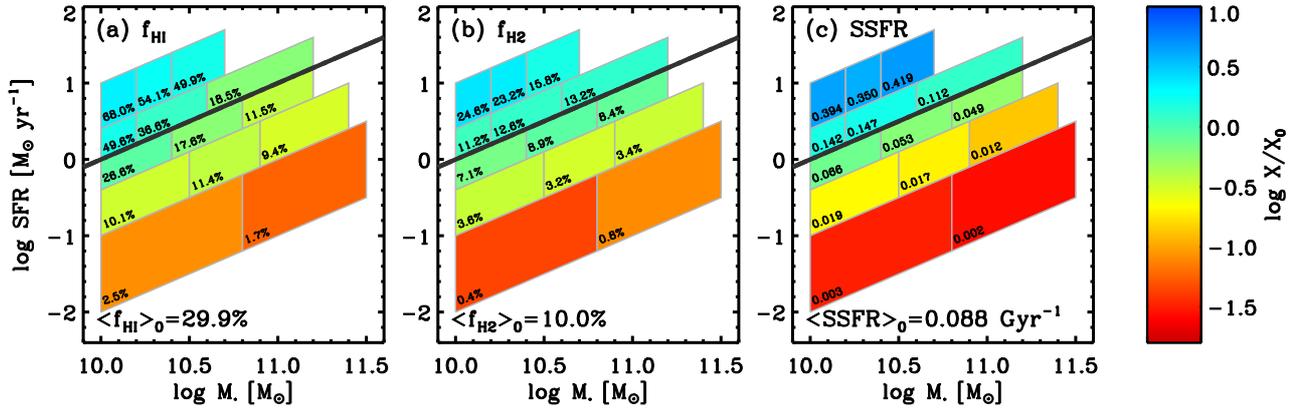}
\caption{Gas properties in the \ms\ plane. Galaxies have been stacked in 14 regions defined by lines of constant SSFR and \mstar\ to produce mean atomic (panel a) and molecular (panel b) gas factions throughout the \ms\ plane. The gas fractions normalised to their value at the fixed characteristic SSFR of 0.1 Gyr$^{-1}$ (thick black line) have been used to color-code the different boxes, meaning that the color scale between all panels can be directly compared. Panel (c) shows as a comparison the mean SSFR in each of the bins; had either $f_{HI}$ or $f_{H2}$ alone been responsible for driving the SSFR of galaxies, either of these panels would have matched this benchmark. \label{gasMS}}
\end{figure*}

\begin{table*}
\centering
\caption{Mean gas fractions in the SFR-\mstar\ plane \label{MStab}}
\begin{tabular}{lcccc}
\hline
$\Delta \log($SSFR) & {$\log M_{\ast}/M_{\odot}$} & {$\log$ SFR$/M_{\ast}$} & {$f_{HI}$} & $f_{H2}$ \\
\hline 
$[ 0.4, 1.0]$ & $[10.0,10.2]$ & $ -9.40\pm0.022$ & $0.680\pm0.146$ & $0.246\pm0.027 $ \\
 & $[10.2,10.4]$ & $ -9.46\pm0.021$ & $0.541\pm0.074$ & $0.232\pm0.020 $ \\
 & $[10.4,10.7]$ & $ -9.38\pm0.028$ & $0.499\pm0.172$ & $0.158\pm0.012 $ \\
\hline
$[ 0.0, 0.4]$ & $[10.0,10.2]$ & $ -9.85\pm0.012$ & $0.496\pm0.032$ & $0.112\pm0.008 $ \\
 & $[10.2,10.6]$ & $ -9.83\pm0.016$ & $0.366\pm0.029$ & $0.126\pm0.008 $ \\
 & $[10.6,11.2]$ & $ -9.95\pm0.013$ & $0.185\pm0.018$ & $0.132\pm0.009 $ \\
\hline
$[-0.4, 0.0]$ & $[10.0,10.4]$ & $-10.18\pm0.014$ & $0.266\pm0.020$ & $0.071\pm0.005 $ \\
 & $[10.4,10.8]$ & $-10.27\pm0.014$ & $0.176\pm0.011$ & $0.089\pm0.007 $ \\
 & $[10.8,11.2]$ & $-10.31\pm0.035$ & $0.115\pm0.012$ & $0.084\pm0.006 $ \\
\hline
$[-1.0,-0.4]$ & $[10.0,10.5]$ & $-10.73\pm0.028$ & $0.101\pm0.007$ & $0.036\pm0.004 $ \\
 & $[10.5,10.9]$ & $-10.77\pm0.019$ & $0.114\pm0.008$ & $0.032\pm0.002 $ \\
 & $[10.9,11.4]$ & $-10.93\pm0.015$ & $0.094\pm0.008$ & $0.034\pm0.003 $ \\
\hline
$[-2.0,-1.0]$ & $[10.0,10.8]$ & $-11.55\pm0.016$ & $0.025\pm0.002$ & $0.004\pm0.001 $ \\
 & $[10.8,11.5]$ & $-11.63\pm0.017$ & $0.017\pm0.001$ & $0.008\pm0.001 $ \\
   \hline
 \textbf{[-0.4, 0.4]} & \textbf{[10.0,11.5]} & \textbf{-10.06} & \textbf{0.299} & \textbf{0.100}  \\
\hline
\end{tabular}
\end{table*}

Because not all the galaxies have a detection of the CO and/or of the HI line, we cannot gain a full understanding of gas across the \ms\ plane simply by using the individually measured line fluxes.  Instead, we use a stacking technique to include both detections and non-detections in meaningful mean values in 14 distinct bins in the \ms\ parameter space.  The bins are outlined in Figure \ref{figsample}.  They are based on the definition of a characteristic timescale for star formation in the local Universe of 0.1 Gyr$^{-1}$.  This is the specific star formation rate of a main sequence galaxy of $10^{10}$\msun\ based on the definition of the MS from \citet{peng10} for a SDSS-selected sample.   We define five bins of SSFR based on the offset from this characteristic timescale, which are then further divided in three stellar mass intervals (two for the lowest SSFR bin).  

To derive mean molecular gas fractions in each bin, all individual spectra are stacked irrespective of whether individually they have a detectable CO emission line or not.  The stacking is done in ``gas fraction units", meaning that all multiplicative factors required to convert an observed CO line flux into a \fgas\ value are applied to each spectrum prior to stacking.  This includes all the standard factors from the \citet{solomon97} prescription, but also \xco, \mstar$^{-1}$, and a weight to ensure that the sample is fully representative of its unbiased parent sample.   The equivalent operation is performed to obtain a mean value of \fhi\ in each bin.  The full details of the stacking methodology are given in \citet{fabello11a} and \citet{saintonge12}.  The results of the stacking experiment in the different SSFR and \mstar\ bins are illustrated in Figure \ref{gasMS} and summarized in Table \ref{MStab}.   The rightmost panel of Figure \ref{gasMS} shows the mean SSFR in each bin, normalized to the characteristic value of 0.1 Gyr$^{-1}$ as represented by the color bar.  We are going to use this as a benchmark to see whether or not we can explain the position of galaxies in the \ms\ plane based on their gas contents. 

First, we look at the atomic gas mass fraction, $f_{\rm HI}=M_{\rm HI}/M_{\ast}$, in the left panel.  Again, the color-coding is done by normalizing to the value along the characteristic SSFR line such that the same color bar applies.  If the HI contents of galaxies alone was responsible for setting their SSFR, then panels a and c of Figure \ref{gasMS} would look identical.   While the overall structure is similar, there is an additional trend for \fhi\ to decrease with \mstar\ at fixed SSFR.  We can perform a more quantitative analysis to determine the direction of variation of \fhi\ in the \ms\ plane.   We use a standard $\chi^2$ minimisation technique to fit a plane to the three-dimensional space defined by \mstar, SFR and \fhi\ using the 14 stacked data points and obtain:
\begin{eqnarray}
 \log f_{HI} &=& 8.57 + 0.588 \log {\rm SFR}  - 0.902 \log M_{\ast}      \nonumber \\
 	&=& 8.57 + 0.588 \log {\rm SSFR}  - 0.314 \log M_{\ast}
\label{fHIeq}
\end{eqnarray}
The sub-linear slope in this relation with SSFR indicates that variations in \fhi\ alone cannot explain the range of SSFR values observed across the $z=0$ \ms\ plane. This can be seen clearly in Fig. \ref{gasMS}a as a lack of dynamic range in the color scale, compared to the reference SSFR plot.  

In the middle panel of Figure \ref{gasMS}, we present the variations of the molecular gas mass fraction, $f_{\rm H_2}=M_{\rm H_2}/M_{\ast}$.  This comes closer to reproducing the SSFR variations (right panel), as unlike in the case of \fhi\ there are no trends with \mstar\ at fixed SSFR.  The figure however still lacks some dynamic range; \fgas\ variations account for most of the variations of SSFR but not completely.  This point was already made in \citet{saintonge12}, where it was shown that a slowly varying star formation efficiency is also required to explain the high values of SSFR in starburst galaxies, and the low values measured in bulge-dominated galaxies.  A similar effect has been observed for galaxies at $z=1-2$ \citep{genzel15}.  We determine the best fitting plane in the \mstar, SFR and \fgas\ data space and find: 
\begin{eqnarray}
 \log f_{H2} &=& 6.02   + 0.704 \log {\rm SFR}  - 0.704 \log M_{\ast}      \nonumber \\
 	&=& 6.02   + 0.704 \log {\rm SSFR}.
\label{fH2eq}
\end{eqnarray}
Both qualitative and quantitative inspection reveal that \fgas\ varies with SSFR with no residual dependence on stellar mass, however once again the sub-linear slope indicates that variations in \fgas\ alone are not enough the explain the full range of SSFR values observed across the \ms\ plane.  The additional quantity that varies in this plane and which accounts for the rest of the dependence of SSFR variations is star formation efficiency, SFE$=$SFR$/M_{H2}$.  It has indeed been shown in previous COLD GASS work that SFE is not constant across the local galaxy population, nor is it a step function with a given value for all normal star-forming galaxies and a higher value for starburst. Rather, star formation efficiency varies smoothly as a function of SSFR \citep{COLDGASS2,saintonge12,huang14,genzel15}.

\subsection{Atomic and molecular gas along the main sequence}

\begin{figure}
\includegraphics[width=\columnwidth]{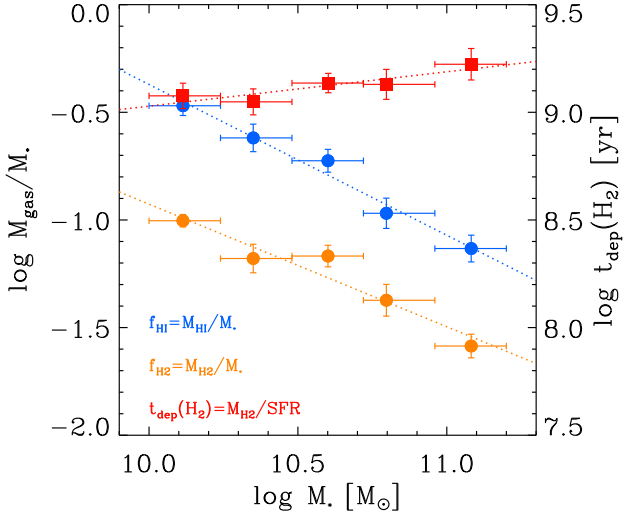}
\caption{Atomic and molecular gas fractions and molecular gas depletion timescale along the main sequence. Galaxies included in this figure are those located within $\pm0.4$dex of the main sequence, which we have parametrised here as eq. \ref{MSeq} by tracking the locus of star-forming galaxies as a function of \mstar. \label{HIH2} }
\end{figure}

\begin{table}
\centering
\caption{Mean gas fractions along the main sequence \label{alongMStab}}
\begin{tabular}{cccc}
\hline
$\log M_{\ast}/M_{\odot}$ & {$f_{HI}$} & {$f_{H2}$}  &{$t_{dep}(H_2)$ [Gyr]} \\
\hline
$[10.00, 10.24]$ & 0.34$\pm$0.04 & 0.10$\pm$0.01 & 1.19$\pm$0.16\\
$[10.24, 10.48]$ & 0.24$\pm$0.04 & 0.07$\pm$0.01 & 1.12$\pm$0.16\\
$[10.48, 10.72]$ & 0.19$\pm$0.02 & 0.07$\pm$0.01 & 1.37$\pm$0.14\\
$[10.72, 10.96]$ & 0.11$\pm$0.02 & 0.04$\pm$0.01 & 1.35$\pm$0.22\\
$[10.96, 11.20]$ & 0.07$\pm$0.01 & 0.03$\pm$0.00 & 1.67$\pm$0.28\\
\hline
\end{tabular}
\end{table}

The ``main sequence" (MS) is loosely defined as the location of star-forming galaxies in the \ms\ plane.  It it often parametrised as a linear relation between $\log {\rm M}_{\ast}$ and $\log {\rm SFR}$, with a slope in the range of 0.5 to 1.0 and a scatter about this relation of $\sim0.3$ dex \citep[e.g.][]{noeske07,pannella09,rodighiero10}.  Another option is to define the MS as the ridge tracing the locus of the bulk of the star-forming galaxies in the \ms\ plane.  For stellar masses lower than $\sim10^{10.5}$\msun\ there is but little difference between the two definitions.  However as shown in Figure \ref{figsample}, in the regime of massive galaxies, there is a clear flattening of the relation obtained by following the locus of the star-forming galaxies (orange line) compared with a constant characteristic SSFR (bold black line).  This flattening has been observed both at low and high redshifts \citep[e.g.][]{karim11,whitaker12}. 

The analysis done in Figure \ref{gasMS} at fixed SSFR helps to understand the link between gas contents and position in the \ms\ plane, but does not indicate why galaxies preferentially populate certain regions on the plane (the MS and the red cloud), or why the MS has a specific shape and scatter.   To address these questions, we use the second approach described above and define a non-linear MS.  We extract all galaxies from the SDSS DR7 with $0.01<z<0.05$ and \mstar$>10^8$\msun, and divide them in stellar mass bins of width 0.15 dex.  In each \mstar\ bin, we fit a Gaussian to all the individual SFR measurements; the central position of this Gaussian is taken to be the characteristic SFR of star-forming galaxies for this given \mstar.  Finally, we fit these characteristic SFRs as a function of \mstar\ with a third order polynomial to obtain our main sequence:
\begin{equation}
\log {\rm SFR} = -2.332 x + 0.4156 x^2 - 0.01828 x^3,
\label{MSeq}
\end{equation}
where $x=\log(M_{\ast}/M_{\odot})$. This relation is shown in Fig. \ref{figsample} over the mass interval \mstar$>10^{10}$ \msun\ where the GASS/COLD GASS sample is located.  We identify as main sequence galaxies those at any given mass with $| \log ({\rm SFR/SFR_{MS}}) |<0.4$.  

This definition in hand, we can now study the properties of galaxies along the main sequence.  In Figure \ref{HIH2}, the mean atomic and molecular gas fractions of MS galaxies are shown as a function of stellar mass.  Although the detection rate is high for MS galaxies (84\% and 89\% for CO and HI, respectively), the mean values are derived through stacking to account for the few non-detections.   Both the atomic and molecular gas mass fractions decline steadily with stellar mass, indicating that the flattening of the MS at $\log M_{\ast}/M_{\odot}>10.0$ is due to the gradual decrease of the gas contents with increasing stellar mass.  Since the depletion timescale along the MS is roughly constant, only increasing very slightly with \mstar\ from $\sim1.2$ to $\sim1.5$ Gyr as shown in Fig. \ref{HIH2}, the decreasing gas fractions explain why the massive galaxies cannot sustain star formation at the same characteristic timescale of 0.1 Gyr$^{-1}$ as in lower mass galaxies.  The molecular-to-atomic ratio ($R_{mol}\equiv M_{H_2}/M_{HI}$) for the main sequence galaxies is $\sim 0.3$ independently of stellar mass, although there is tentative evidence that this ratio increases at lower stellar masses.  Further work on similar galaxy samples with \mstar$<10^{10}$\msun\ will be necessary to confirm this.  

The fact that both the molecular-to-atomic ratio and the depletion timescale are close to constant for galaxies on the MS reinforces the conclusion that the flattening of the MS at \mstar$>10^{10}$\msun\ is not caused by bottlenecks in the conversion of atomic to molecular gas or in the star formation process, but rather by the overall reduction of the gas reservoirs of these galaxies.

\subsection{Morphology and stellar ages along the main sequence}

\begin{figure*}
\includegraphics[width=\textwidth]{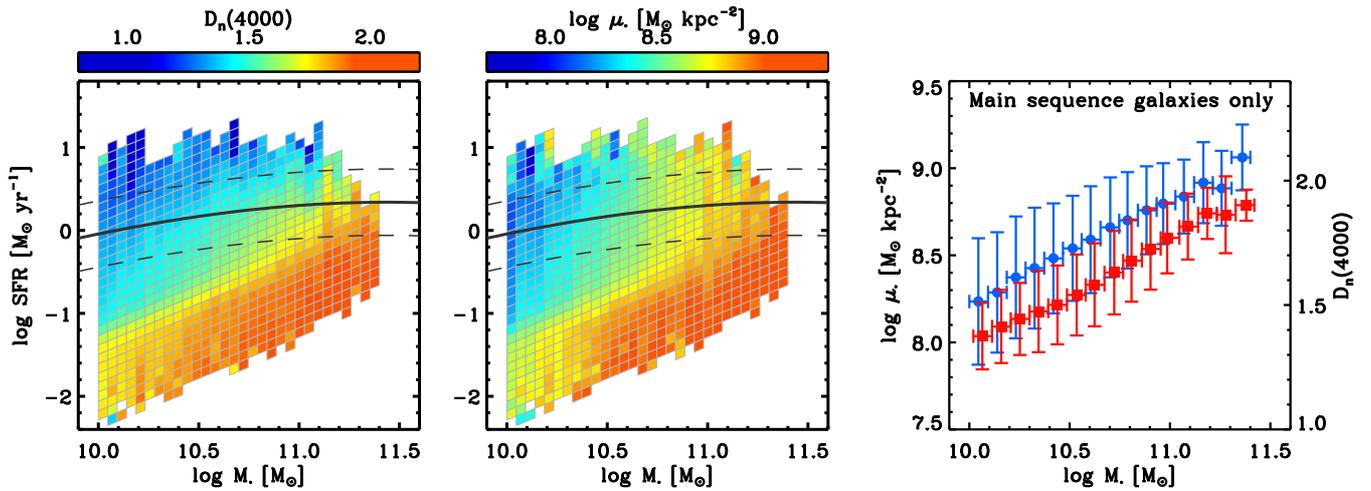}
\caption{Mean $D_n(4000)$ index (left) and stellar mass surface density, \must\ (middle), in the \ms\ plane. The solid black line shows the position of the main sequence of star-forming galaxies as parametrised by Eq. \ref{MSeq}, with dashed line showing the region within $\pm0.4$ dex of the relation. The right panel shows the mean values of \must\ (red squares) and $D_n(4000)$ (blue circles) along the main sequence. \label{morphMS}}
\end{figure*}

A large number of physical properties have been reported to vary between galaxies located on and above the main sequence, including dust temperature, IR$/$UV ratio, star formation efficiency, stellar population properties and morphology \citep[e.g.][]{elbaz11,wuyts11,magnellisaintonge,nordon13}.  Above, we have shown that for massive galaxies it is important to make a distinction between variation that happens at fixed SSFR and along the main sequence (e.g. \fgas\ is constant at fixed SSFR but decreases along the MS, while the reverse is true for$R_{mol}$).  Therefore to understand the changes in gas fraction along the MS, we revisit some of these observations for our SDSS sample. We choose the D$_n$(4000) index to track the age of the stellar population in the central region of the galaxies,  and stellar mass surface density, $\mu_{\ast}$, as an indicator of morphology (the same results are obtained if using another proxy such as the concentration index).  

In Figure \ref{morphMS}, we show how these two quantities vary across the \ms\ plane, making use of the entire SDSS parent sample of 12006 galaxies from which the GASS/COLD GASS samples are extracted.  As in Figure \ref{gasMS}, the diagonal lines defining the bins are lines of constant SSFR, while the solid line shows the position of the MS (eq. \ref{MSeq}).   Both quantities, but in particular the D$_n$(4000) index, vary mostly with SSFR rather than with \mstar.  Since the MS is significantly flatter than a line of constant SSFR, the mean values of both \must\ and D$_n$(4000) increase monotonically along the MS as shown in the right panel of Figure \ref{morphMS}.   Therefore, the reduction of the atomic and molecular gas fractions of MS galaxies as \mstar\ increases goes hand in hand with bulge growth and the ageing of the stellar population in the central regions of these galaxies.

\subsection{The molecular ratio and star formation quenching}

\begin{figure}
\includegraphics[width=\columnwidth]{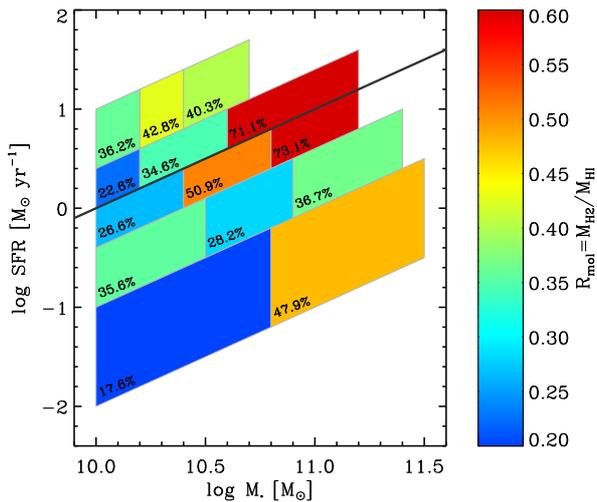}
\caption{Variations of the molecular-to-atomic ratio in the \ms\ plane. The values of $R_{mol}$ are derived from the stacking results summarised in Figure \ref{gasMS}.  \label{rmol} }
\end{figure}

Because we are seeing the reduction of the gas fraction in MS galaxies as \mstar\ increases, we ask whether what we are witnessing is the quenching of star formation through a gradual reduction of the gas reservoirs.  Some insights into this question are contained in the balance between the atomic and molecular gas components. In particular, the different behaviours of \fhi\ and \fgas\ observed in Figure \ref{gasMS} (and quantified by Equations \ref{fHIeq} and \ref{fH2eq}) reveal the existence of systematic variations in the molecular-to-atomic ratio, $R_{mol}=M_{H2}/M_{HI}$, throughout the \ms\ plane.  This is shown explicitly in Figure \ref{rmol}, with mean values ranging from 25\% for the galaxies with the lowest masses and SFRs to 70\% for massive galaxies with high SFRs.  To interpret these large variations, it is important to first remember what \rmol\ is measuring here.  The CO and HI line fluxes are obtained with telescope beams of 22\arcsec\ and 3.5\arcmin, respectively.  Aperture corrections are used to infer a total \mh\ value from the observed CO line flux, but as the HI observations are unresolved it is impossible to obtain a measure of \mhi\ over a similar aperture; the Arecibo-measured \mhi\ unavoidably include a contribution from diffuse atomic gas located well outside the optical diameter of the galaxies where most of the molecular gas is contained.  Therefore, our measurement of \rmol\ tells us to some extent about the conversion of atomic to molecular gas within the star-forming discs of galaxies, but much more about the relative importance of those large extended HI reservoirs compared to the amount of star-forming gas.  

Galaxies with the highest SSFRs probed here have on average \rmol$\sim$40\%, slightly over the average of $30\%$ found for the entire COLD GASS sample \citep{COLDGASS1}.  In the other four SSFR slices sampled in Figure \ref{rmol}, there is a trend for increasing \rmol\ with stellar mass.  This is nowhere more striking than for galaxies with $-10.4<\log {\rm SSFR}<-9.6$ where the most massive galaxies have on average \rmol$>70\%$.  Without extended HI reservoirs (such as are typical in lower mass galaxies) or significant gas accretion, these galaxies will not be able to sustain star formation at such high SSFR for more that one depletion timescale ($\sim 1$Gyr) and are therefore likely candidates for galaxies being caught just shortly before being quenched.  This observation is discussed further in  \S \ref{summary}. 

\begin{figure}
\includegraphics[width=\columnwidth]{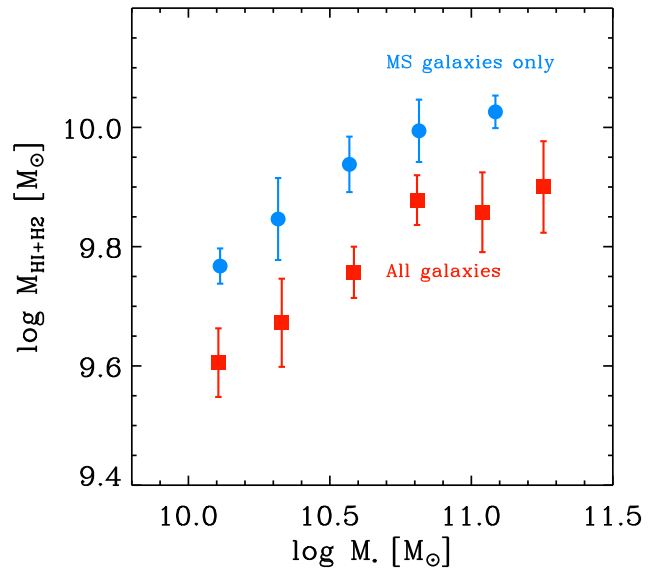}
\caption{The total cold gas contents of main sequence galaxies (blue circles) and the entire sample (red squares) as a function of stellar mass. \label{mtot} }
\end{figure}

A distinction needs to be made between defining quenching in mass normalised, or ``specific", terms (e.g. the reduction of SSFR or \fgas) or in absolute terms (e.g. SFR or \mh). While gas fractions decrease along the MS (Fig. \ref{HIH2}), the absolute values of the gas mass do in fact increase.  Figure \ref{mtot} illustrates how the total cold gas mass of galaxies (\mhi$+$\mh) increases with \mstar. This is true both for MS galaxies separately, and for the entire galaxy population, indicating a need for ongoing accretion of cold gas, even in massive galaxies in the local universe. This accretion has to be significant enough not only to maintain SFR roughly constant for MS galaxies with \mstar$>10^{10}$\msun\ (see Fig. \ref{figsample}), but also to account for the (small yet significant) increase in the total cold gas mass.

\section{Summary of observational results and discussion}
\label{summary}

The objective of this paper was to provide a first quantitative measure of the mean atomic and molecular gas contents of galaxies across the entire \ms\ plane for galaxies with \mstar$>10^{10}$\msun.  This is made possible by the GASS / COLD GASS surveys, which have obtained HI 21cm and CO(1-0) line flux measurements for a large, representative SDSS-selected sample. Even though individual galaxies are not all detected in the HI and/or CO line, we can derive accurate mean gas fractions over the entire \ms\ plane using a spectral stacking technique, owing to the homogeneous observing strategy of the surveys and their well-defined selection functions out of the larger SDSS parent sample.  Building on these strengths, we set out with the following objectives: (1)  to track systematically how gas contents, star formation efficiency and the molecular-to-atomic ratio vary across the \ms\ plane, (2) to understand how, in an average sense, the position of galaxies in the plane are related to their gas contents, (3) to identify regions of the plane where galaxies are out of equilibrium and in the process of quenching, and (4) to follow specifically the gas properties of the objects that form the main sequence of star forming galaxies. 

To address questions (1)-(3) we separated the \ms\ plane in 14 bins defined by lines of constant SSFR and \mstar. Stacking the HI and CO spectra gave us mean atomic and molecular gas fractions in each of these bins (see Figure \ref{gasMS}). We find that \fhi\ can explain to first order the SSFRs of different galaxy populations, although note that \fhi\ has an additional dependence on \mstar\ (eq. \ref{fHIeq}) and cannot alone account for the very high SSFRs of above main sequence galaxies. On the other hand, \fgas\ depends tightly on SSFR with no residual \mstar\ dependence because the molecular-to-atomic ratio is a function of stellar mass, with the most massive galaxies at fixed SSFR having a higher $R_{mol}$.  

Our quantitative analysis of the variations of the molecular gas mass fraction in the \ms\ plane (eq. \ref{fH2eq}) reveals the sub-linear slope of the relation between \fgas\ and SSFR; this implies that star formation efficiency must also be varying with SSFR to account for the full range of values \citep[a result previously reported in a number of studies, including ][]{saintonge12,genzel15}. We therefore conclude that, on average, {\it the position of galaxies in the \ms\ plane are determined by (1) how much cold gas is present as traced by HI, (2) how much of that cold gas is in the molecular phase and available for star formation, and (3) how efficient is the process of converting that molecular gas into stars.}  

We also identify a region of the \ms\ plane ($\log M_{\ast}/M_{\odot}>10.8$ and $-10.4<\log{\rm SSFR}<-9.6$) that is populated by galaxies which are within $\sim1$ Gyr of quenching.  Galaxies in this ``danger zone" have on average very high molecular-to-atomic mass ratios, $R_{mol}>0.7$, more than twice as large as the mean for main-sequence galaxies. In the absence of extended HI envelopes or other sources of accretion to replenish the gas reservoirs, these galaxies will cease actively forming stars and migrate to the red cloud. Galaxies in the ``danger zone" are disc-like but with important bulge components ($<\log \mu_{\ast}>=8.9$), while also having young stellar populations in their central regions ($<D_n(4000)>=1.38$). This is an unusual combination: throughout the rest of the \ms\ plane, there is a strong correlation between bulge-like morphology and older stellar ages (see Fig. \ref{morphMS}). Indeed, a control sample to the galaxies in the ``danger zone" matched on \mstar\ and \must\ has a distribution of values of $D_n(4000)$ with a Kolmogorov-Smirnoff probability $<0.001\%$ of being extracted from the same parent sample, and a mean value of $<D_n(4000)>=1.85\pm0.05$. Our interpretation is that the bulge-dominated galaxies in the ``danger zone" benefit from a mechanism that efficiently drives gas towards their central, high-density regions, explaining the large values of $R_{mol}$ and the high levels of star formation in the central bulge region driving down the $D_n(4000)$ index. We observe that strong stellar bars are a common feature of galaxies in the ``danger zone", and are indeed an example of a mechanism that can trigger inward radial gas motions and therefore increased central gas concentrations and star formation rates \citep[e.g.][]{sakamoto99,sheth05,masters12}. 

We also looked specifically at the properties of galaxies along the main sequence, which importantly we define here not as a linear relation between $\log M_{\ast}$ and $\log {\rm SFR}$, but by following the ridge traced by star-forming galaxies in the \ms\ plane.  We take main sequence galaxies to be those with $| \log ({\rm SFR/SFR_{MS}}) |<0.4$ dex (see Fig. \ref{figsample} and Eq. \ref{MSeq}).  As shown in Figure \ref{HIH2}, the mean molecular and atomic gas fractions of galaxies decline steadily with \mstar\ along the main sequence. In contrast, both the molecular gas depletion timescale, \tdep, and the molecular-to-atomic ratio, $R_{mol}$ are very near constant along the main sequence.  This implies that the reason for the flattening of the MS at \mstar$>10^{10}$\msun\ is the gradual decrease in the total cold gas mass fraction of star-forming galaxies, and not because of a reduction of the conversion of atomic into molecular gas or of the efficiency of star formation. 

Other quantities are well known to vary systematically across the \ms\ plane, including measures of stellar population properties and different morphological indicators such as stellar mass surface density, concentration index, and S\'ersic index. As these quantities appear mostly constant at fixed SSFR \citep[see Fig. \ref{morphMS} and e.g.][]{wuyts11}, the flattening of the MS also means they vary systematically as star-forming galaxies grow more massive. Our analysis of the properties of massive galaxies along the main sequence thus reveals that multiple transformations are occurring: galaxies grow central bulges, the mean age of their stellar populations increases, and their entire cold gas reservoir (atomic and molecular) decreases.  

A significant outstanding question in galaxy evolution studies concerns the mechanisms responsible for quenching, meant as the transition of galaxies from the star forming to the passive population.  A vast number of mechanisms have been explored to explain quenching. These include processes that actively remove gas from galaxies such as feedback from AGN and star formation or stripping \citep[e.g.][]{gunngott72,cicone14,forster14}, to more passive mechanisms that simply prevent galaxies from accreting fresh gas \citep[e.g.][]{keres05,dekel06,peng15}.  Such mechanisms are usually invoked to explain the relatively quick transition of galaxies form the main sequence to the passive population, but our results here suggest that quenching has already started happening for massive galaxies while on the main sequence. The mechanism responsible for this must be able to account for the simultaneous reduction of the gas fractions of massive main sequence galaxies, ageing of their stellar population and growth of their central bulges.

\section*{Acknowledgements}

AS acknowledges the support of the Royal Society through the award of a University Research Fellowship.  BC is the recipient of an Australian Research Council Future Fellowship (FT120100660). This research was supported under the Australian Research Council's Discovery Projects funding scheme (DP130100664 and DP150101734). RG and MPH and the ALFALFA team at Cornell have been  supported by NSF grants AST-0607007 and AST-1107390 and by grants from the Brinson Foundation.

This work is based on observations carried out with the IRAM 30m telescope and the Arecibo Observatory. IRAM is supported by INSU/CNRS (France), MPG (Germany), and IGN (Spain).   The Arecibo Observatory is operated by SRI International under a cooperative agreement with the National Science Foundation (AST-1100968), and in alliance with Ana G. M\'endez-Universidad Metropolitana, and the Universities Space Research Association.  

This publication makes use of data products from the Wide-field Infrared Survey Explorer, which is a joint project of the University of California, Los Angeles, and the Jet Propulsion Laboratory/California Institute of Technology, funded by the National Aeronautics and Space Administration

\bibliographystyle{mnras}

\bsp	
\label{lastpage}
\end{document}